\documentclass[aip,12pt,graphicx]{revtex4}

\usepackage{comment, color, graphicx,amsmath}

\newcommand{\that}{\hat{t}}

\newcommand{\vv}{v}

\newcommand{\ve}{\vv_{\mathrm{e}}}

\newcommand{\fM}{f_{\mathrm{M}}}

\newcommand{\gradp}{\nabla_{\vect{p}}}

\newcommand{\nuee}{\nu_{\mathrm{ee}}}

\newcommand{\vect}[1]{\mbox{\boldmath $#1$}}

\newcommand{\be}{\begin{equation}}
\newcommand{\ee}{\end{equation}}

\newcommand{\mylsim}{\mathrel{\mbox{\raisebox{-1mm}{$\stackrel{<}{\sim}$}}}}

\newcommand{\ypa}{y_{\|}}
\newcommand{\ype}{y_{\perp}}

\newcommand{\gr}{{\tilde\gamma}}

\newcommand{\changed}[1]{{#1}}

\begin{document}


\title{\changed{Numerical calculation of the runaway electron distribution function and associated synchrotron emission}}



\author{Matt Landreman}
\email[]{mattland@umd.edu}
\affiliation{Institute for Research in Electronics and Applied Physics, University of Maryland, College Park, MD, 20742, USA}
\affiliation{Plasma Science and Fusion Center, MIT, Cambridge, MA, 02139, USA}
\author{Adam Stahl}
\author{T\"unde F\"ul\"op}
\affiliation{Department of Applied Physics, Nuclear Engineering,
  Chalmers University of Technology and Euratom-VR Association,
  G\"oteborg,
  Sweden}


\date{\today}

\begin{abstract}

\changed{Synchrotron emission from runaway electrons may be used to diagnose plasma
conditions during a tokamak disruption, but solving this inverse problem
requires rapid simulation of the electron distribution function and associated synchrotron emission
as a function of plasma parameters.
Here we detail a framework for this forward calculation, beginning with an efficient numerical method} for solving the
Fokker-Planck equation in the
presence of an electric field of arbitrary strength.  The approach
is continuum (Eulerian), and we employ a relativistic collision
operator, valid for arbitrary energies.  Both primary and secondary
runaway electron generation are included.  For cases in which primary generation dominates, a
time-independent formulation of the problem is described, requiring
only the solution of a single sparse linear system.
\changed{
In the limit of dominant secondary generation, we present the first numerical
verification of an analytic model for the distribution function.
The numerical electron distribution function
in the presence of both primary and secondary generation is then used for calculating the
synchrotron emission spectrum of the runaways.
It is found that the average synchrotron spectra emitted from realistic distribution functions
are not well approximated by the emission of a single electron
at the maximum energy.}

\end{abstract}


\maketitle 


\section{Introduction}
Due to the decrease in the Coulomb collision cross section
with velocity, charged particles in an electric
field can ``run away'' to high energies.  In tokamaks, the resulting energetic
particles can damage plasma-facing components and are expected to be a
significant danger in the upcoming ITER experiment. Electrons are
typically the species for which runaway is most significant \cite{Dreicer,Connor}, but runaway ions \cite{Furth} and positrons \cite{Helander,Fulop} can also
be produced.  Relatively large electric fields are required for
runaway production, and in tokamaks these can arise during disruptions
or in sawtooth events.  Understanding of runaway electrons and their
generation and mitigation is essential to planning future large
experiments such as ITER.

\changed{Runaway electrons emit measurable synchrotron radiation, which can potentially be used
to diagnose the distribution function, thereby constraining the physical parameters
in the plasma. The runaway distribution function and associated synchrotron emission
depend on the time histories of the local electric field $E$, temperature $T$, average ion charge $Z$, and density $n$.
To infer these quantities (and the uncertainty in these quantities) inside a disrupting
plasma using the synchrotron emission, it is necessary to run many simulations of the runaway process,
scanning the various physical parameters.  To make such a scan practical, computational efficiency is important.
}

\changed{To this end, in this work we demonstrate
a framework for rapid computation of the runaway distribution function and associated
synchrotron emission for given plasma parameters.
The distribution function is computed using a new numerical tool named CODE
(COllisional Distribution of Electrons).
Physically, the distribution function is determined by}
a balance between acceleration in the electric field and collisions
with both electrons and ions.  The calculation in CODE is fully relativistic,
using a collision operator valid for both low and high velocities
\cite{papp} and it includes both primary and secondary runaway
electron generation.  If primary runaway electron generation
dominates, CODE can be used in both time-dependent and
time-independent modes. The latter mode of operation, in which a
long-time quasi-equilibrium distribution function is calculated, is
extremely fast in that it is necessary only to solve a single sparse
linear system.
\changed{Due to its speed and simplicity, CODE is highly suitable for
coupling within larger more expensive calculations.
Besides the inverse problem of determining plasma parameters from synchrotron
emission, other such applications could include the
study of instabilities driven by the anisotropy of the electron
distribution function, and comprehensive modeling of disruptions.}

\changed{
Other numerical methods for computing the distribution function of runaways
have been demonstrated previously, using a range of algorithms.
Particle methods follow the trajectories of individual marker electrons.
Deterministic particle calculations \cite{MartinSolis} can give insight into the system
behavior but cannot calculate the distribution function, since diffusion is absent.
Collisional diffusion may be included by making random adjustments to particles' velocities,
an approach which has been used in codes such as ASCOT\cite{ASCOT} and ARENA\cite{ARENA}.
For a given level of numerical uncertainty (noise or discretization error),
we will demonstrate that CODE is more than 6 orders of magnitude faster
than a particle code on the same computer.
Other continuum codes developed to model energetic electrons include BANDIT\cite{BANDIT}, CQL3D\cite{CQL1, CQL2} and LUKE \cite{LUKE03, LUKE08}.
These sophisticated codes were originally developed to model RF heating and current drive,
and contain many features not required for the calculations we consider. For example,
CQL3D contains $\sim 90,000$ lines of code and LUKE contains $\sim 118,000$ lines, whereas CODE contains $<1,200$ lines (including comments).
While future more elaborate modeling may require the additional features of
a code like CQL3D or LUKE, for the applications we consider, we find it useful to have the nimble and dedicated tool CODE.
For calculations of non-Maxwellian distribution functions in the context of RF heating,
an adjoint method \cite{Karney86} can be a useful technique for efficient solution of linear inhomogeneous
kinetic equations. However, the kinetic equation we will consider is nonlinear (if avalanching is included) and homogeneous,
so the adjoint method is not applicable.
}

\changed{
In several previous studies, a single particle with a representative momentum and pitch-angle is used as
an approximation for the entire runaway distribution
\cite{Jaspers,Yu} when computing the synchrotron emission.  In this paper, we present a computation of the
synchrotron radiation spectrum of a runaway distribution in various
cases. By showing the difference between these spectra and those based on single particle emission we
demonstrate the importance of taking into account the entire distribution.}

The remainder of the paper is organized as follows. In
Sec.~\ref{sec:keq} we present the kinetic equation and the collision
operator used. Section \ref{sec:num} details the discretization scheme
and calculation of the primary runaway production rate, with typical
results shown in section \ref{sec:primaryResults}.  The avalanche
source term and its implementation are described in
Sec.~\ref{sec:avalanche}.
\changed{In this section we also demonstrate agreement with an analytic
model for the distribution function \cite{Fulop2006}.
Computation of the synchrotron emission spectrum from the distribution function is detailed in
Sec.~\ref{sec:synchrotron}, and comparisons to single-particle emission are given.}  We conclude in Sec.~\ref{sec:concl}.

\section{Kinetic Equation and Normalizations}
\label{sec:keq}
We begin with the kinetic equation
\begin{equation}
\frac{\partial f}{\partial t}-eE\vect{b}\cdot\gradp f = C\{f\} + S.
\label{eq:ke}
\end{equation}
Here, $-e$ is the electron charge, \changed{$E$ is the component of
the electric field along the magnetic field,} $\vect{b}=\vect{B}/B$ is a unit vector
along the magnetic field, $\gradp$ is the gradient in the
space of relativistic momentum $\vect{p} = \gamma m \vect{v}$,
$\gamma=1/\sqrt{1-v^2/c^2}$, $v=|\vect{v}|$ is the speed, $m$ is the
electron rest mass, $c$ is the speed of light, $C$ is the electron
collision operator, and $S$ represents any sources.  All quantities
refer to electrons unless noted otherwise.
\changed{Equation (\ref{eq:ke}) is the large-aspect-ratio limit
of the bounce- and gyro-averaged Fokker-Planck equation (e.q. (2) in (\cite{rosenbluth})).
Particle trapping effects are neglected, which is reasonable since runaway
beams are typically localized close to the magnetic axis. }
We may write $\vect{b}\cdot\gradp f$ in (\ref{eq:ke}) in
terms of scalar variables using
\begin{equation}
\vect{b}\cdot\gradp f = \xi \frac{\partial f}{\partial p} + \frac{1-\xi^2}{p} \frac{\partial f}{\partial \xi}
\end{equation}
where $p=|\vect{p}|$, and
$\xi = \vect{p}\cdot\vect{b}/p$ is the cosine of the pitch angle relative to the magnetic field.
The distribution function is defined such that
the density $n$ is given by $n = \int d^3p\; f$,
so $f$ has dimensions of $(\mathrm{length}\times\mathrm{momentum})^{-3}$, and we assume the distribution function for small momentum to be
approximately the Maxwellian
$\fM = n \pi^{-3/2} (m \ve)^{-3} \exp(-y^2)$ where $\ve=\sqrt{2T/m}$ is the thermal
speed, and $y=p/(m\ve) = \gamma v/\ve$ is the normalized momentum.

We use the collision operator from Appendix B of Ref.~\cite{papp}.
This operator is constructed to match the usual nonrelativistic test-particle operator in the limit of $v \ll c$, and in the relativistic limit it reduces to the operator from Appendix A of Ref.~\cite{Sandquist}.
The collision operator is
\begin{equation}
C\{f\} = \frac{1}{p^2} \frac{\partial}{\partial p} p^2 \left[ C_A\frac{\partial f}{\partial p}+C_Ff\right] + \frac{C_B}{p^2}\frac{\partial}{\partial\xi}(1-\xi^2)\frac{\partial f}{\partial \xi}
\label{eq:C}
\end{equation}
where
\begin{eqnarray}
C_A &=& \frac{\Gamma}{v} \Psi(x), \\
C_B &=& \frac{\Gamma}{2v} \left[ Z + \phi(x) - \Psi(x) + \frac{\delta^4 x^2}{2}\right], \\
C_F &=& \frac{\Gamma}{T} \Psi(x),
\end{eqnarray}
$\delta = \ve/c$, $x = v/\ve = y/\sqrt{1+\delta^2 y^2}$, $Z$ is the effective ion charge,
\begin{equation}
\Gamma = 4\pi n e^4 \ln\Lambda = (3\sqrt{\pi}/4)\nuee\ve^3 m^2
\end{equation}
is identical to the $\Gamma$ defined in Refs.~\cite{papp,Sandquist,Karney},
$\nuee = 4\sqrt{2\pi}e^4 n \ln\Lambda/(3\sqrt{m}T^{3/2})$ is the usual
Braginskii electron collision frequency, $\phi(x) =
2\pi^{-1/2}\int_0^x \exp(-s^2)\,ds$ is the error function, and
\begin{equation}
\Psi(x) = \frac{1}{2x^2}\left[ \phi(x)-x \frac{d\phi}{dx}\right]
\end{equation}
is the Chandrasekhar function.  In the nonrelativistic limit
$\delta\to 0$, then $y\to x$, and (\ref{eq:C}) reduces to the usual
Fokker-Planck test-particle electron collision operator.

The collision operator (\ref{eq:C}) is approximate in several
ways. First, it originates from the Fokker-Planck approximation in
which small-angle collisions dominate, which is related to an
expansion in $\ln\Lambda \gg 1$.  Consequently, the infrequent
collisions with large momentum exchange are ignored, so the secondary
avalanche process is not included at this stage, but will be addressed
later in Sec.~\ref{sec:avalanche}. Also, the modifications to the Rosenbluth
potentials associated with the high-energy electrons are neglected,
i.e. collisions with high-energy field particles are ignored.

The kinetic equation is normalized by multiplying through with $m^3
\ve^3\pi^{3/2}/(\nuee n)$, and defining the normalized distribution
function
\begin{equation}
F = (\pi^{3/2} m^3 \ve^3/n) f
\end{equation}
so that $F \to 1$ at $p \to 0$. We also introduce a normalized electric field
\begin{equation}
\hat{E} = - e E/(m \ve \nuee)
\end{equation}
which, up to a factor of order unity, is $E$ normalized by the Dreicer
field.  The normalized time is $\that = \nuee t$ and the normalized
source is $\hat S = S m^3 \ve^3\pi^{3/2}/(\nuee n)$.  We thereby
obtain the dimensionless equation
\begin{eqnarray}
\frac{\partial F}{\partial \that}
+\hat{E} \xi \frac{\partial F}{\partial y} + \hat{E} \frac{1-\xi^2}{y} \frac{\partial F}{\partial y}
-\frac{3\sqrt{\pi}}{4}\frac{1}{y^2} \frac{\partial}{\partial y} y^2 \left[ \frac{\Psi(x)}{x} \frac{\partial F}{\partial y} + 2\Psi(x) F \right] && \label{eq:normalizedKE}\\
-\frac{3\sqrt{\pi}}{4}\frac{1}{2xy^2}
\left[Z + \phi(x) - \Psi(x) + \frac{\delta^4 x^2}{2}\right]
\frac{\partial}{\partial\xi}(1-\xi^2)\frac{\partial F}{\partial\xi}
&=&\hat{S} \nonumber .
\end{eqnarray}
Notice that this equation has the form of a linear inhomogeneous 3D partial
differential equation:
\begin{equation}
\frac{\partial F}{\partial \that} + M F = \hat{S}
\end{equation}
for a linear time-independent differential operator $M$.  If a
time-independent equilibrium solution exists, it will be given by $F =
M^{-1} \hat{S}$.

Since both the electric field acceleration term and the collision
operator in the kinetic equation (\ref{eq:ke}) have the form of a
divergence of a flux in velocity space, the total number of particles
is constant in time in the absence of a source: $(d/dt)\int d^3p\; f =
\int d^3p\; S$.  However, runaway electrons are constantly gaining
energy, so without a source at small $p$ and a sink at large $p$, no
time-independent distribution function will exist. From another perspective, a nonzero source is necessary to find a
nonzero equilibrium solution of (\ref{eq:normalizedKE}), because when
\mbox{$\hat{S}=0$}, (\ref{eq:normalizedKE}) with $\partial/\partial\that =0$
is a homogeneous equation with homogeneous boundary conditions.  (The
boundary conditions are that $F$ be regular at $y=0$, $\xi=-1$, and
$\xi=1$, and that $f\to 0$ as $y\to\infty$.)  Thus, the solution of
the time-independent problem $F = M^{-1}\hat{S}$ for $\hat{S}=0$ would
be $F=0$.

To find a
solution, we must either consider a time-dependent problem or include
a nonzero $S$.  In reality, spatial transport can give rise to both
sources and sinks, and a sink exists at high energy due to
radiation. When included, secondary runaway generation (considered in
Sec. \ref{sec:avalanche}) also introduces a source.
To avoid the added complexity of these sinks and sources and simultaneously
avoid the intricacies of time dependence, when restricting ourselves to primary generation we may formulate a time-independent problem as follows. We take $\hat S = \alpha
e^{-y^2}$ for some constant $\alpha$, representing a thermal source of
particles. Equation (\ref{eq:normalizedKE}) for $\partial/\partial\that
=0$ may be divided through by $\alpha$ and solved for the unknown
$F/\alpha$.  Then $\alpha$ may be determined by the requirement
$F(p=0)=1$, and $F$ is then obtained by multiplying the solution
$F/\alpha$ by this $\alpha$.

The constant $\alpha$ represents the rate at which particles must be
replenished at low energy to balance their flux in velocity space to
high energy.  Therefore, $\alpha$ is the rate of runaway production.
As we do not introduce a sink at high energies, $F$ will have a
divergent integral over velocity space.

CODE can also be run in time-dependent mode.  Once the velocity space
coordinates and the operator $M$ are discretized, any implicit or
explicit scheme for advancing a system of ordinary differential
equations (forward or backward Euler, Runge-Kutta, trapezoid rule, etc.) may be applied to the time coordinate. (Results shown in this paper are computed using the trapezoid rule.)  Due to the diffusive nature of $M$, numerical stability favors implicit time-advance schemes.

\section{Discretization}
\label{sec:num}

We first expand $F$ in Legendre polynomials $P_L(\xi)$:
\begin{equation}
F(y,\xi) = \sum_{L=0}^\infty F_L(y) P_L(\xi).
\end{equation}
Then the operation
\begin{equation}
\frac{2L+1}{2}\int_{-1}^1 P_L(\xi) ( \;\;\;\cdot\;\;\;)d\xi
\label{eq:discretizer}
\end{equation}
is applied to the kinetic equation.
Using the identities in the appendix, we obtain
\begin{eqnarray}
\frac{\partial F_L}{\partial \that}
+ \sum_{\ell=0}^{\infty}  \left\{
\hat{E} \left[
\frac{L+1}{2L+3}\delta_{L+1,\ell} + \frac{L}{2L-1}\delta_{L-1,\ell}\right] \frac{\partial}{\partial y}
\right. && \label{eq:LegendreKE} \\
+\frac{\hat{E}}{y}
\left[ \frac{(L+1)(L+2)}{2L+3}\delta_{L+1,\ell} - \frac{(L-1)L}{2L-1}\delta_{L-1,\ell}\right]
&&\nonumber \\
-\frac{3\sqrt{\pi}}{4} \frac{\Psi(x)}{x} \delta_{L,\ell} \frac{\partial^2}{\partial y^2}
-\frac{3\sqrt{\pi}}{2} \left[ \frac{2\Psi(x)}{y} + \frac{dx}{dy} \frac{d\Psi}{dx}\right] \delta_{L,\ell}
&& \nonumber \\
-\frac{3\sqrt{\pi}}{4} \left[ \frac{1}{x}\frac{dx}{dy}\frac{d\Psi}{dx}
+\frac{2\Psi(x)}{xy} - \frac{\Psi(x)}{x^2} \frac{dx}{dy} + 2\Psi(x)\right]
\delta_{L,\ell} \frac{\partial}{\partial y}&& \nonumber \\
\left.
+\frac{3\sqrt{\pi}}{8 x y^2} \left[ Z+ \phi(x) - \Psi(x) + \frac{\delta^4 x^2}{2}\right]L(L+1)\delta_{L,\ell}
\right\}F_\ell &=&\hat{S}_L \nonumber
\end{eqnarray}
where
$dx/dy = (1+\delta^2 y^2)^{-3/2}$,
$d\Psi/dx = 2\pi^{-1/2}e^{-x^2}-(2/x)\Psi(x)$,
and
$\hat{S}_L = (2L+1)\, 2^{-1} \int_{-1}^1 \hat{S}\; d\xi$
is the appropriate Legendre mode of $\hat{S}(y,\xi) = \sum_{L=0}^{\infty} \hat{S}_L(y) P_L(\xi) $.
Note that the collision operator is diagonal in the $L$ index, and the electric field acceleration
term is tridiagonal in $L$.

It is useful to examine the $L=0$ case of (\ref{eq:LegendreKE}),
which corresponds to (half) the integral of the kinetic equation over $\xi$:
\begin{equation}
\frac{\partial F_0}{\partial \that}
- \frac{1}{y^2} \frac{\partial}{\partial y} \left[
-y^2 \frac{\hat E}{3}F_1 + \frac{3\sqrt{\pi}}{4} y^2
\left\{ \frac{\Psi(x)}{x} \frac{\partial F_0}{\partial y} + 2\Psi(x)F_0\right\}\right]
= \hat{S}_0.
\label{eq:Lzero}
\end{equation}
Applying $4\pi^{-1/2} \int_{y_b}^\infty dy\; y^2(\;\;\;\cdot\;\;\;)$ for some boundary
value $y_b$, and assuming the source is negligible in this region,
we obtain
\begin{equation}
\frac{1}{\nuee n} \frac{d n_r}{d\that}
= -\frac{4}{\sqrt{\pi}}
\left[
-y^2 \frac{\hat E}{3}F_1 + \frac{3\sqrt{\pi}}{4} y^2
\left\{ \frac{\Psi(x)}{x} \frac{\partial F_0}{\partial y} + 2\Psi(x)F_0\right\}
\right]_{y = y_b}
\label{eq:dnrdt}
\end{equation}
where $n_r$ is the number of runaways,
meaning the number of electrons with $y>y_b$,
so that
$n_r = \int_{y>y_b} d^3p \;f = 2\pi \int_{m\ve y_b}^\infty dp\; p^2\int_{-1}^1d\xi\; f$.
The runaway rate calculated from
(\ref{eq:dnrdt}) should be independent of
$y_b$ in steady state (as long as $y_b$ is in a region of $\hat{S}_0=0$),
which can be seen
by applying $\int_{y_{b1}}^{y_{b2}} dy\; y^2 (\;\;\;\cdot\;\;\;)$
to (\ref{eq:Lzero}).
We find in practice it is far better to compute
the runaway production rate using (\ref{eq:dnrdt})
than from the source magnitude $\alpha$,
since the latter is more sensitive to the various numerical resolution parameters.

To discretize the equation in $y$, we can apply fourth-order finite
differences on a uniform grid.
Alternatively, for greater numerical efficiency,
a coordinate transformation can be applied so grid points are
spaced further apart at high energies.
The $y$ coordinate is cut off at some
finite maximum value $y_{max}$.
The appropriate boundary conditions
at $y=0$ are $d F_0/dy=0$ and $F_L=0$ for $L>0$.  For the boundary at
large $y$, we impose $F_L=0$ for all $L$.  This boundary condition
creates some unphysical grid-scale oscillation at large $y$, which may
be eliminated by adding an artificial
\changed{diffusion
$c_1 y^{-2} (\partial/\partial y) y^2 \exp(-[y-y_{max}]/c_2) \partial/\partial y$
localized near
$y_{max}$ to the linear operator. Suitable values for the constants are $c_1=0.01$ and $c_2=0.1$.
This term effectively represents a sink for particles, which must be included in the time-independent
approach due to the particle source at thermal energies.
Since this diffusion term is exponentially small away from $y_{max}$,
the distribution function is very insensitive to the details of the ad-hoc term except very near
$y_{max}$.
All results shown hereafter are very well converged
with respect to doubling the domain size $y_{max}$, indicating
the results are insensitive to the details of the diffusion term.
}

\section{Results for primary runaway electron generation}
\label{sec:primaryResults}

\begin{figure}
\includegraphics[scale=0.99]{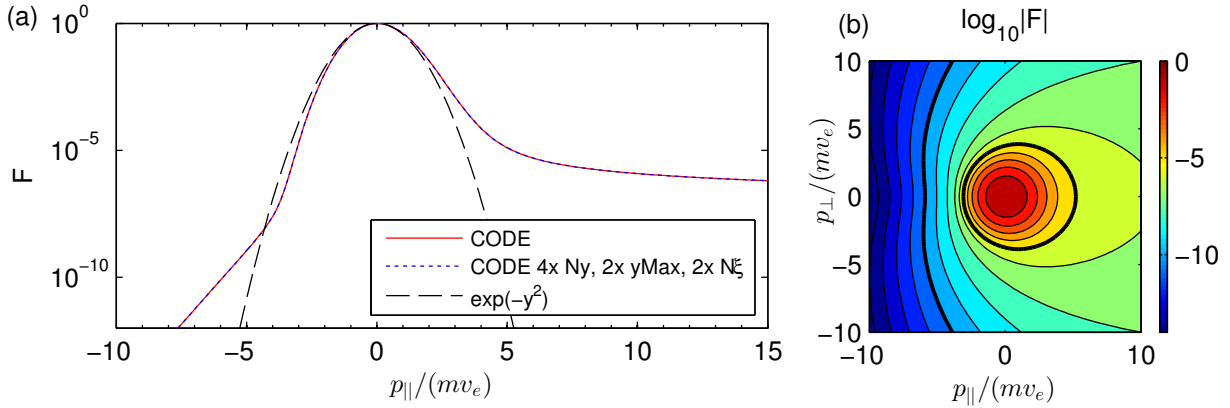}
\caption{(Color online)
Typical results of CODE, obtained for $\delta=0.1$,  $\hat{E} = 0.1$, and $Z=1$.
(a) Normalized distribution function $F$ for $p_\bot=0$.
Results are plotted for two different sets of numerical parameters
($\{N_y=300, y_{max}=20, N_\xi=20\}$ and $\{N_y=1200, y_{max}=40, N_\xi=40\}$).
The results overlap completely, demonstrating excellent
convergence. A Maxwellian is also plotted for comparison.
(b) Contours of $F$ at values $10^z$ for integer $z$.  Bold contours indicate $F=10^{-5}$ and $10^{-10}$.
\label{fig:typicalCODE}}
\end{figure}

Figure \ref{fig:typicalCODE} shows typical results from a
time-independent CODE computation.  To verify convergence, we may
double $N_\xi$ (the number of Legendre modes), double $N_y$ (the
number of grid points in $y$), and double the maximum $y$ ($y_{max})$
at fixed $y$ grid resolution (which requires doubling $N_y$ again.)
As shown by the overlap of the solid red and dashed blue curves in
Fig.~\ref{fig:typicalCODE}a, excellent convergence is achieved for
the parameters used here.  \changed{Increasing the ad-hoc diffusion
magnitude $c_1$ by a factor of 10 for the parameters of the red
curve causes a relative change in the runaway rate (computed using
(\ref{eq:dnrdt}) for $y_b=10$) of $|dn_r/d\that(c_1=0.1) -
dn_r/d\that(c_1=0.01)|/ [dn_r/d\that(c_1=0.01)] < 10^{-9}$,
demonstrating the results are highly insensitive to this diffusion
term.}  As expected, the distribution function is increased in the
direction opposite to the electric field ($p_{||}>0$).  While the
distribution function is reduced in the direction parallel to the
electric field ($p_{||}<0$) for $y < 5$, $F$ is actually slightly
increased for $y > 5$ due to pitch-angle scattering of the high-energy
tail electrons, \changed{an effect also seen in Fokker-Planck
simulations of RF current drive \cite{Karney79}.  The pitch-angle
scattering term can be artificially suppressed} in CODE, in which
case $F$ is reduced in the direction parallel to the electric field
for all $y$.

\begin{figure}
\includegraphics{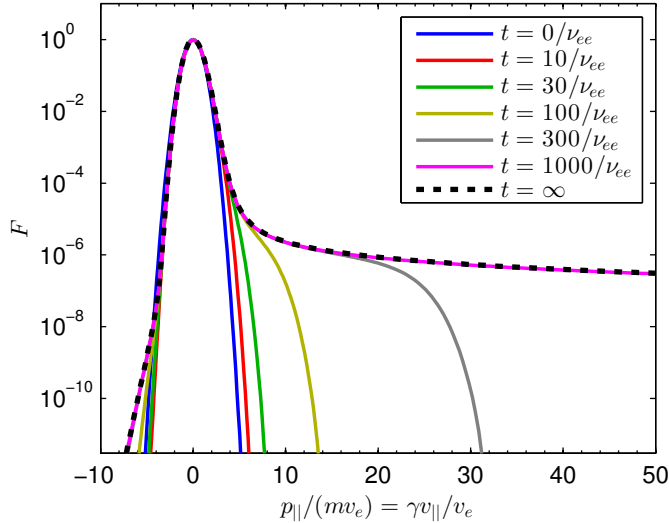}
\caption{(Color online)
The distribution function from time-dependent CODE at various times.
At $t = 1000 / \nuee$, the distribution function is indistinguishable
from the solution obtained using the time-independent scheme ($t=\infty$)
over the momentum range shown.
\label{fig:timeDependence}}
\end{figure}

Figure \ref{fig:timeDependence} compares the distribution functions
obtained from the time-independent and time-dependent approaches.
At sufficiently long times, the time-dependent version produces results that
are indistinguishable from the time-independent version.

\begin{figure}
\includegraphics{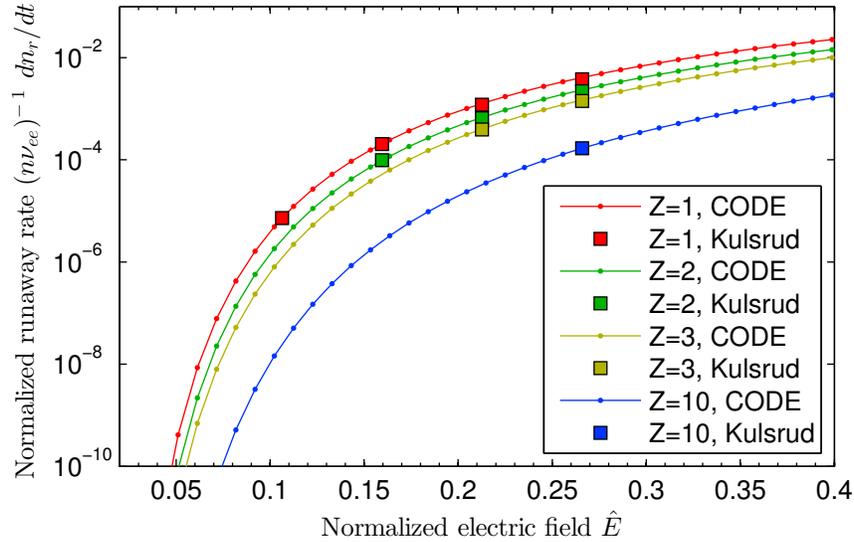}
\caption{(Color online)
Benchmark of CODE in the nonrelativistic limit $\delta\to 0$ against
data in Table 1 of Ref.~\cite{Kulsrud}.
\label{fig:KulsrudBenchmark}}
\end{figure}

For comparison with previously published results, we show in Figure
\ref{fig:KulsrudBenchmark} results by Kulsrud et al \cite{Kulsrud}, who
considered only the nonrelativistic case $\delta\to 0$.  The agreement
with CODE is exceptional.  The runaway production rate in CODE is
computed using (\ref{eq:dnrdt}) for $y_b=10$. (Any value of $y_b>5$
gives indistinguishable results.)  Ref.~\cite{Kulsrud} uses a
different normalized electric field $E_K$ which is related to
$\hat{E}$ by $E_K = 2(3\sqrt{\pi})^{-1}\hat{E}$, and in
Ref.~\cite{Kulsrud} the runaway rate is also normalized by a
different collision frequency $\nu_K = 3\sqrt{\pi/2}\,\nuee$.  It should
also be noted that the Kulsrud computations are time-dependent, with a
simulation run until the flux in velocity space reaches an approximate
steady state.  Each CODE point shown in figure
\ref{fig:KulsrudBenchmark} took approximately 0.08s on a single Dell
Precision laptop with Intel Core i7-2860 2.50 GHz CPU and 16 GB
memory, running in MATLAB.  Faster results could surely be obtained
using a lower-level language.

\changed{ To emphasize the speed of CODE, we have directly compared it
to the ARENA code \cite{ARENA} for computing the runaway rate using
the parameters considered in \cite{Kulsrud}.  ARENA is a Monte Carlo
code written in Fortran 90 and designed specifically to compute the
runaway distribution function and runaway rate. Detailed description
of the current version of ARENA is given in
Refs.~\cite{Ryden,Csepany}.  Both codes were run on a single thread
on the same computer with an Intel Xeon 2.0 GHz processor. ARENA required 49,550
seconds to reproduce the left square point in figure
\ref{fig:KulsrudBenchmark}, and 5,942 seconds to reproduce the
top-right square point. 50,000 particles were required for
reasonable convergence.  For comparison, at a similar level of convergence,
time-independent CODE required 0.00106 s and 0.000696 s for the two
respective points, and time-dependent CODE required 0.0307 s and
0.00082 s respectively. Thus, for these parameters, both
time-independent and time-dependent CODE require less than
$1.5\times 10^{-7}$ as many cpu-hours as ARENA for the same
hardware.}

\section{Secondary runaway electron generation}
\label{sec:avalanche}

In the previous sections we used the Fokker-Planck collision
operator, which includes ``distant'' (large impact parameter) collisions but not ``close'' (small impact parameter) collisions
in which a large fraction of energy and momentum are transferred
between the colliding particles.
Close collisions are infrequent compared to distant collisions,
and are therefore neglected in the Fokker-Planck operator. However, close
collisions may still have a significant effect on runaway generation,
since the density of runaways is typically much smaller than the density of
thermal electrons which may be accelerated in a close collision.
The production of runaways through close collisions is known
as secondary production, or as avalanche production since it may occur with exponential growth.
To simulate secondary generation of energetic electrons,
we use a source term derived in \cite{rosenbluth}, starting from the M\o ller scattering cross-section in the $w \gg 1$ limit,
with \mbox{$w=p/(mc) = \delta y$} a normalized momentum. In this limit,
the trajectories of the primary electrons are not
much deflected by the collisions. The source then takes the form
\begin{equation}
S=\frac{n_r}{4 \pi \tau \ln\Lambda}
\delta(\xi - \xi_2) \frac{1}{w^2}\frac{\partial}{\partial w}\left(\frac{1}{1-\sqrt{1+w^2}}\right),
\label{eq:secondarySource}
\end{equation}
where $1/\tau=4\pi n e^4 \ln{\Lambda}/( m^2 c^3)$ is the collision
frequency for relativistic electrons, $n_r$ is the density of the fast electrons and
$\xi_2 = w/(1+\sqrt{1+w^2})$ is the cosine of the pitch angle at which
the runaway is born.  (Our Eq.~(\ref{eq:secondarySource}) differs by a
factor $m^3 c^3$ compared to the source in Ref.~\cite{rosenbluth}
since we normalize our distribution function as $n=\int d^3p\; f$
instead of $n=\int d^3 w\; f$.  There is also a factor of $2\pi$
difference due to the different normalization of the distribution
function.)

Due to the approximations used to derive $S$, care must be taken in
several regards.  First, to define $n_r$ in
(\ref{eq:secondarySource}), it is not clear where to draw the dividing
line in velocity space between runaways and non-runaways.  One
possible strategy for defining $n_r$ is to compute the separatrix in
velocity space between trajectories that will have bounded and
unbounded energy in the absence of diffusion, and to define the
runaway density as the integral of $f$ over the latter region
\cite{Smith}.  This approach may somewhat overestimate the true
avalanche rate, since it neglects the fact that some time must elapse
between an electron entering the runaway region and the electron
gaining sufficient energy to cause secondary generation.  As most
runaways have $\xi \approx 1$, we may approximate the separatrix by
setting $dw/dt=0$ where $dw/dt = eE/(mc) - (1+1/w^2)/\tau$ defines the
trajectory of a particle with $\xi=1$, neglecting diffusion in
momentum and pitch angle. The runaway region is therefore $w>w_c$
where $w_c = \left[ (E/E_c)-1 \right]^{-1/2}$ and \changed{$E_c = m c
  / (e \tau)$} is the critical field, and so we take \mbox{$n_r = 2\pi
  m^3 c^3 \int_{-1}^1 d\xi \int_{w_c}^{\infty} dw\; w^2 f$.}  (We
cannot define $n_r$ by the time integral of (\ref{eq:dnrdt}), since
(\ref{eq:dnrdt}) is no longer valid when $S$ is nonzero away from $p
\approx 0$.)  A second deficiency of (\ref{eq:secondarySource}) is
that $S$ is singular at $w \to 0$, so the source must be cut off below
some threshold momentum.
Following Ref.~\cite{CQL2}, we choose
the cutoff to be $w_c$.
Neither of the cutoffs discussed here would be necessary if a less approximate
source term than (\ref{eq:secondarySource}) were used,
but derivation of such an operator is beyond the scope of this paper.

Normalizing and applying
(\ref{eq:discretizer}) as we did previously for the other terms in the
kinetic equation, the source included in CODE becomes
\begin{equation}
\hat{S}_L = \frac{n_r}{n} \frac{3\pi\delta^5}{16 \ln\Lambda} \frac{2L+1}{2} P_L(\xi_2)
\frac{1}{(1-\sqrt{1+w^2})^2 \sqrt{1+w^2} y}.
\end{equation}
When secondary generation is included,
CODE must be run in time-dependent mode.

\begin{figure}
\includegraphics[scale=0.5]{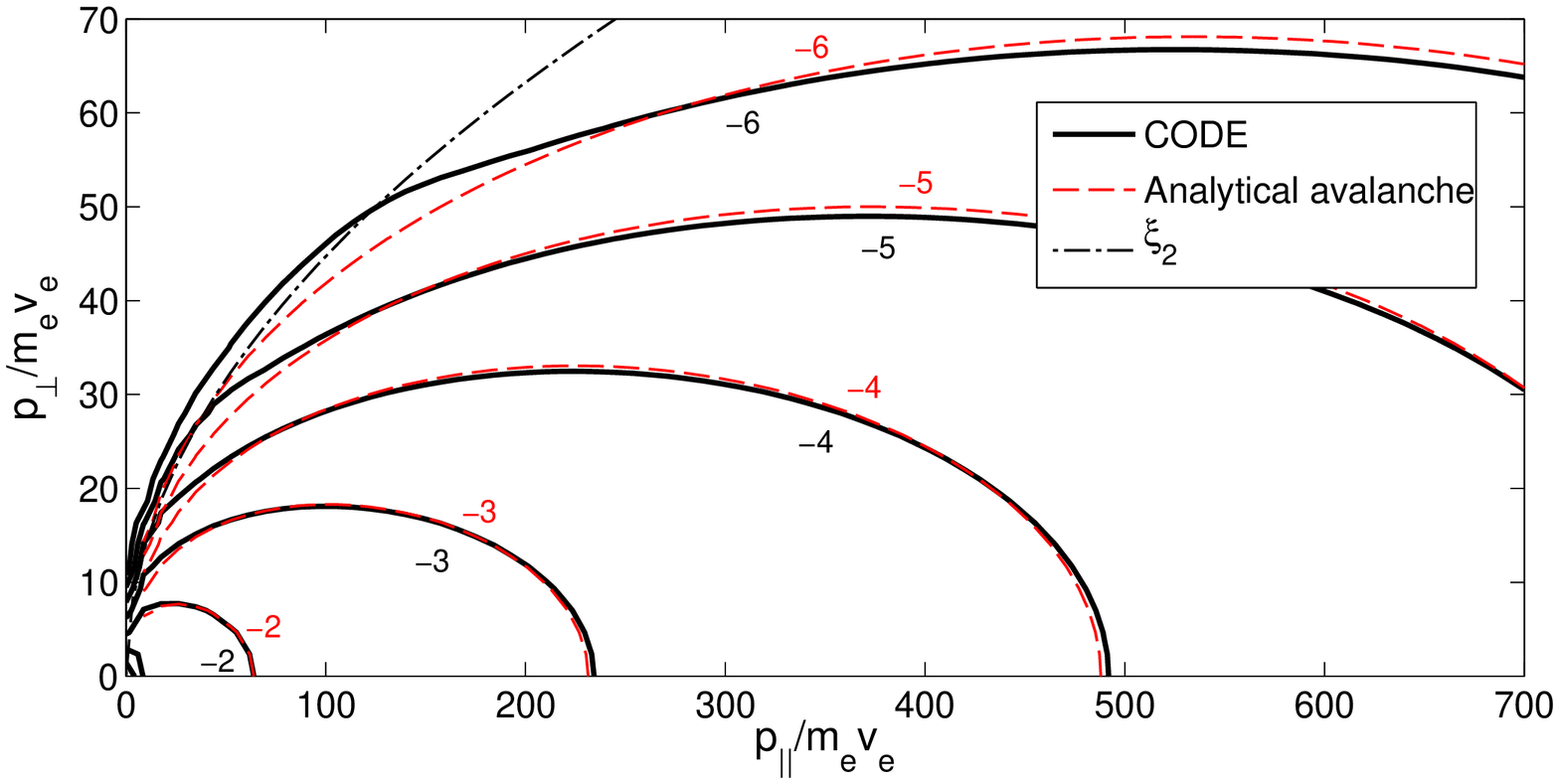}
\includegraphics[scale=0.5]{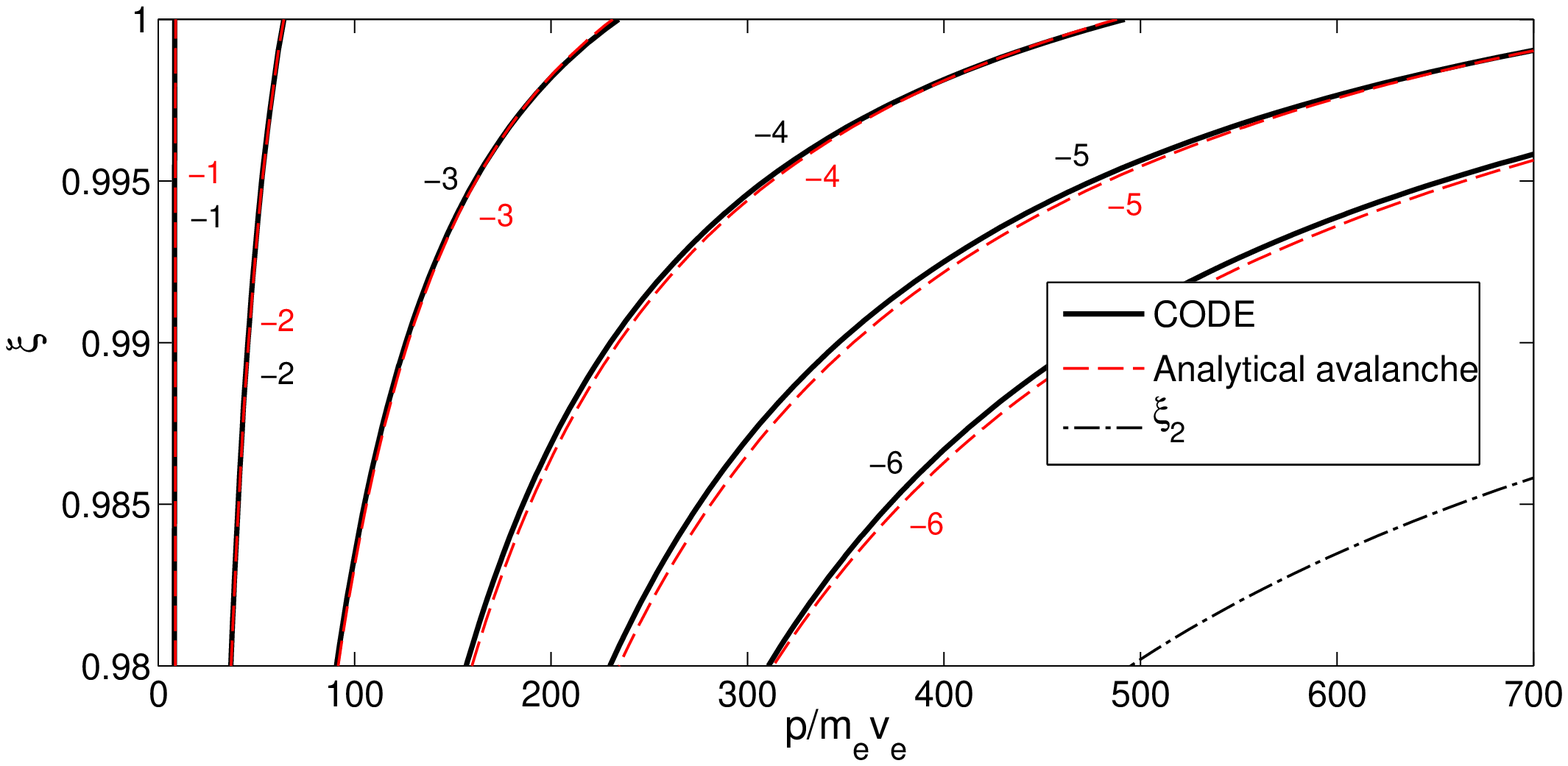}
\caption{(Color online) Contour plots of the long-time distribution
function from CODE (shown in two different coordinate systems),
obtained for $E/E_c=40$ $(\hat{E}=0.532)$, $Z=3$, $\delta=0.1$ and
$t=5000/\nuee$. Results are plotted for the numerical parameters
$N_y=1500$, $y_{max}=1500$ and $N_\xi=100$, with time step
$dt=10/\nuee$. The analytical distribution in \eqref{eq:anal_ava}-\eqref{eq:anal_ava_normalization} for the same
physical parameters is also plotted for comparison, together with
part of the curve where avalanche runaways are created ($\xi = \xi_2$).
\label{fig:contours7}}
\end{figure}

\begin{figure}
\includegraphics[scale=0.5]{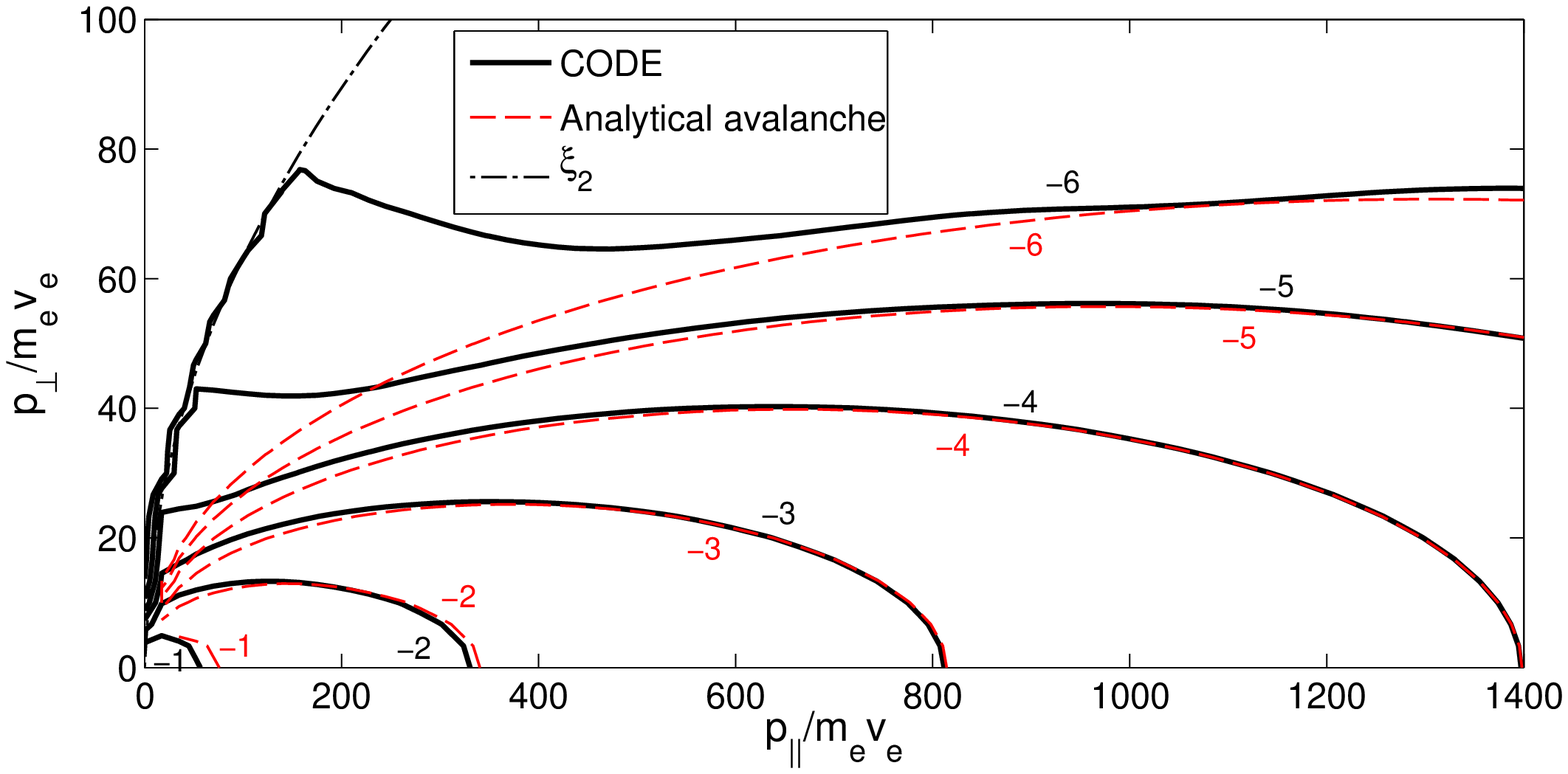}
\includegraphics[scale=0.5]{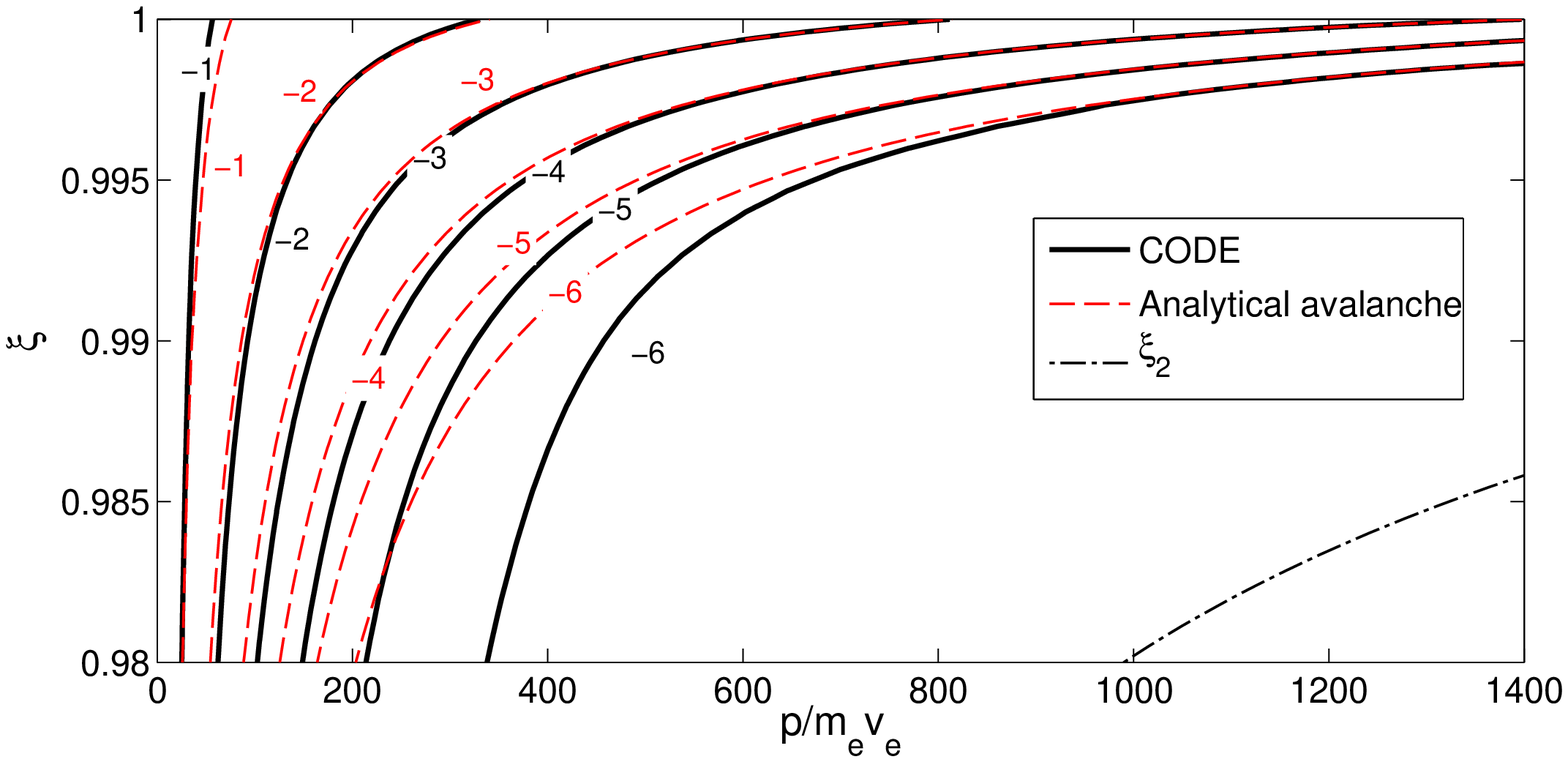}
\caption{(Color online) Contour plots of the long-time distribution function
from CODE (shown in two different coordinate systems), obtained for
$E/E_c=100$ $(\hat{E}=0.332)$, $Z=1$, $\delta=0.05$ and $t=6000/\nuee$.  Results are
plotted for the numerical parameters $N_y=1500$, $y_{max}=3000$ and
$N_\xi=180$, with time step $dt=25/\nuee$.  The analytical
distribution in \eqref{eq:anal_ava}-\eqref{eq:anal_ava_normalization} is also plotted for comparison, together with
part of the curve where avalanche runaways are created ($\xi = \xi_2$).
\label{fig:contours8}}
\end{figure}

To benchmark the numerical solution of the kinetic equation including the above source term by CODE,
we use the approximate analytical expression for the avalanche distribution function derived in Section II of Ref.~\cite{Fulop2006}:
\begin{equation}
f_{aa}(w_{\parallel}, w_{\perp}) = \frac{k}{w_{||}} \exp\left(
\gr t - \frac{\gr\tau}{E/E_c-1}w_{||} - \left[ \frac{E/E_c-1}{Z+1}\right] \frac{w_{\perp}^2}{2w_{||}}
\right)
\label{eq:anal_ava}
\end{equation}
where
$k$ is a constant.
The quantity $\gr$ is the growth rate $\gr=(1/f) \partial f/\partial t$,
which must be independent of both time and velocity for (\ref{eq:anal_ava}) to be valid.
Equation (\ref{eq:anal_ava}) is also valid only where $p_{||} \gg p_{\perp}$
and in regions of momentum space where $S$ is negligible.  (This restriction is not a major one since $S=0$
everywhere except on the $\xi=\xi_2$ curve.)
If most of the runaway distribution function is accurately
described by (\ref{eq:anal_ava}), then we
may approximate $n_r \approx \int d^3 p\; f_{aa} = 2\pi m^3 c^3 \int_{-\infty}^{\infty} dw_{\parallel} \int_0^{\infty} dw_{\perp}\; w_{\perp} f_{aa}$,
giving $\gr = (1/n_r) dn_r/dt$ and
\begin{equation}
k= n_r e^{- \gr t} \frac{\tau}{2\pi m^3 c^3 (1+Z)}
\label{eq:anal_ava_normalization}
\end{equation}
where $n_r e^{-\gr t}$ is constant.
(Equation \eqref{eq:anal_ava_normalization} may be inaccurate in some situations
even if \eqref{eq:anal_ava} is accurate in part of velocity-space, because
\eqref{eq:anal_ava_normalization} requires \eqref{eq:anal_ava} to apply in \emph{all} of velocity-space.)
 Figures \ref{fig:contours7} and
\ref{fig:contours8} show comparisons between distributions from CODE
and \eqref{eq:anal_ava}-\eqref{eq:anal_ava_normalization} for two different sets of parameters.
More precisely, the quantity plotted in figures \ref{fig:contours7}-\ref{fig:contours8} is $\log_{10} (m^3c^3 f / n_r)$.
To generate the figures, CODE is run for a sufficiently long time that $(1/f) \partial f/\partial t$
becomes approximately constant. The resulting numerical value of $(1/n_r)dn_r/dt$ is then used as $\gr$ when evaluating
\eqref{eq:anal_ava}-\eqref{eq:anal_ava_normalization}.
For a cleaner comparison between CODE and analytic theory in these figures, we minimize primary generation in CODE
in these runs by initializing $f$ to 0 instead of to a Maxwellian.
For both sets of physical parameters,
the agreement between CODE and \eqref{eq:anal_ava} is excellent in the region where
agreement is expected: where $p_{||} \gg p_{\perp}$ and away from the curve
$\xi=\xi_2$.

\section{Synchrotron emission}
\label{sec:synchrotron}
\changed{Using the distribution functions calculated with CODE, we now proceed
to compute the spectrum of emitted synchrotron radiation.}
Due to the energy dependence
of the emitted synchrotron power, the emission from runaways
completely dominates that of the thermal particles. The emission also
depends strongly on the pitch-angle of the particle. In a cylindrical
plasma geometry, the emitted synchrotron power per wavelength at
wavelength $\lambda$ from a single highly energetic particle is given
by \cite{Bekefi}
\begin{equation}
\quad \mathcal{P}(\gamma,\gamma_\|,\lambda)=\frac{4\pi}{\sqrt{3}}\,\frac{ce^{2}}{\lambda^{3}\gamma^{2}}\int_{\lambda_{c}/\lambda}^{\infty}\!\!\! K_{5/3}(l)\,dl\ ,\quad \label{eq:Bekefi}
\end{equation}
where the two-dimensional momentum of the particle is determined by
$\gamma$ and \mbox{$\gamma_\|=1/\sqrt{1-v_\|^2/c^2}$}, $K_\nu(x)$ is a
modified Bessel function of the second kind, and
\begin{equation}
\lambda_{c}=\frac{4\pi}{3}\frac{mc^2\gamma_{\parallel}}{eB\gamma^{2}}\ ,
\end{equation}
where $B$ is the magnetic field strength.

Using CODE we will demonstrate that the synchrotron radiation spectrum
from the entire runaway distribution is substantially different from
the spectrum obtained from a single particle approximation. By transforming to the
more suitable coordinates $w$ and $\xi$, related to $\gamma$ and
$\gamma_\|$ through $\gamma^2 = 1+w^2$ and
$\gamma_\|^2=(1-w^2\xi^2/(1+w^2))^{-1}$, and integrating
\eqref{eq:Bekefi} over the runaway region $R$ in momentum space, we
obtain the total synchrotron emission from the runaway
distribution. Normalizing to $n_r$, we find that the average emitted
power per runaway particle at a wavelength $\lambda$ is given by
\begin{equation}
P(\lambda)=\frac{2\pi}{n_{r}}\int_{R}f(w,\xi)\, \mathcal{P}(w,\xi,\lambda)\, w^{2}\mbox{d}w\,\mbox{d}\xi\ .\label{eq:Plambda}
\end{equation}
\changed{
Up to a factor $e c A$, where $A$ is the area of the runaway beam, normalization by $n_r$ is equivalent
to normalization by the runaway current, since the emitting particles
all move with velocity $\approx c$.}

\begin{figure}
\includegraphics[scale=0.5]{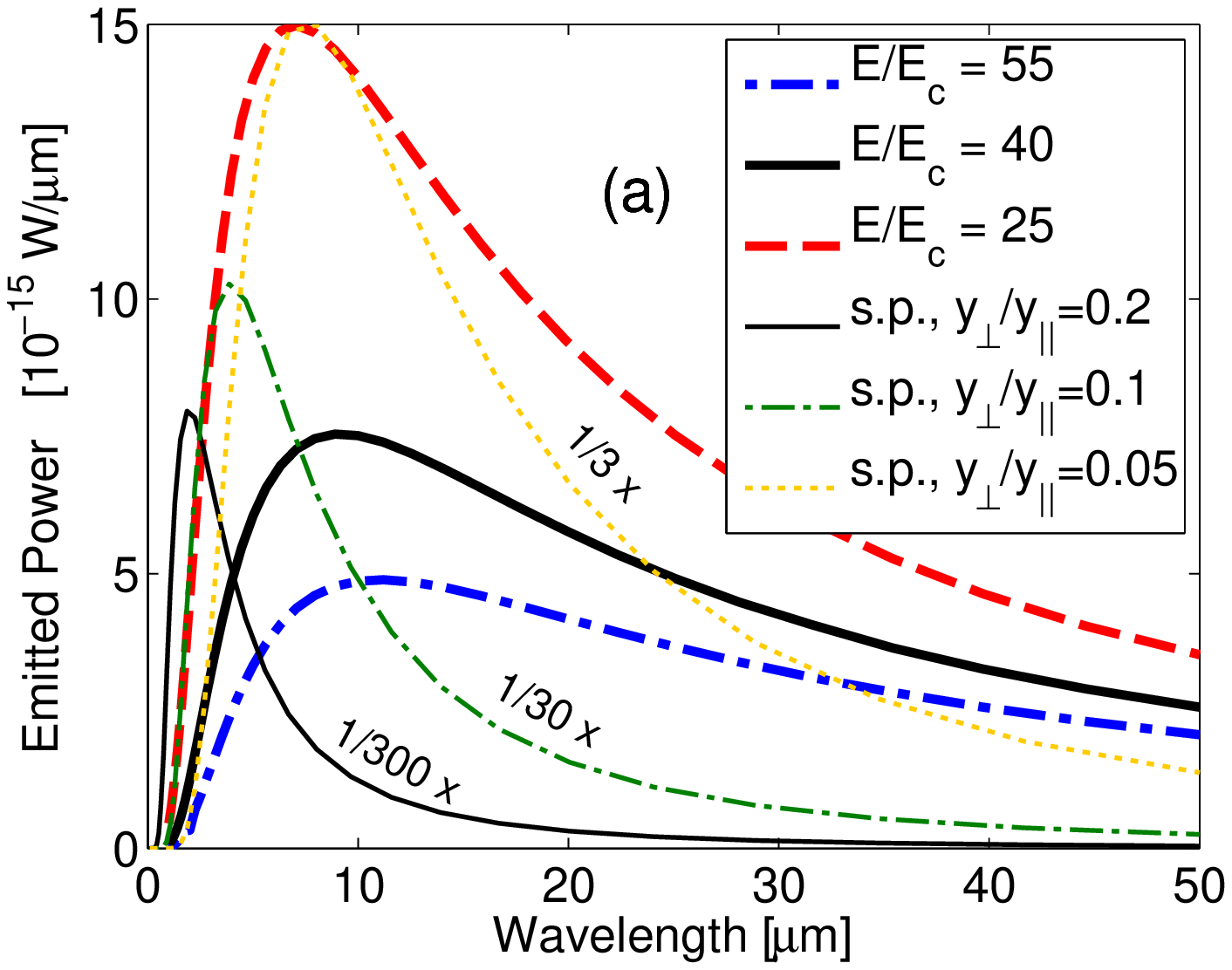}
\includegraphics[scale=0.5]{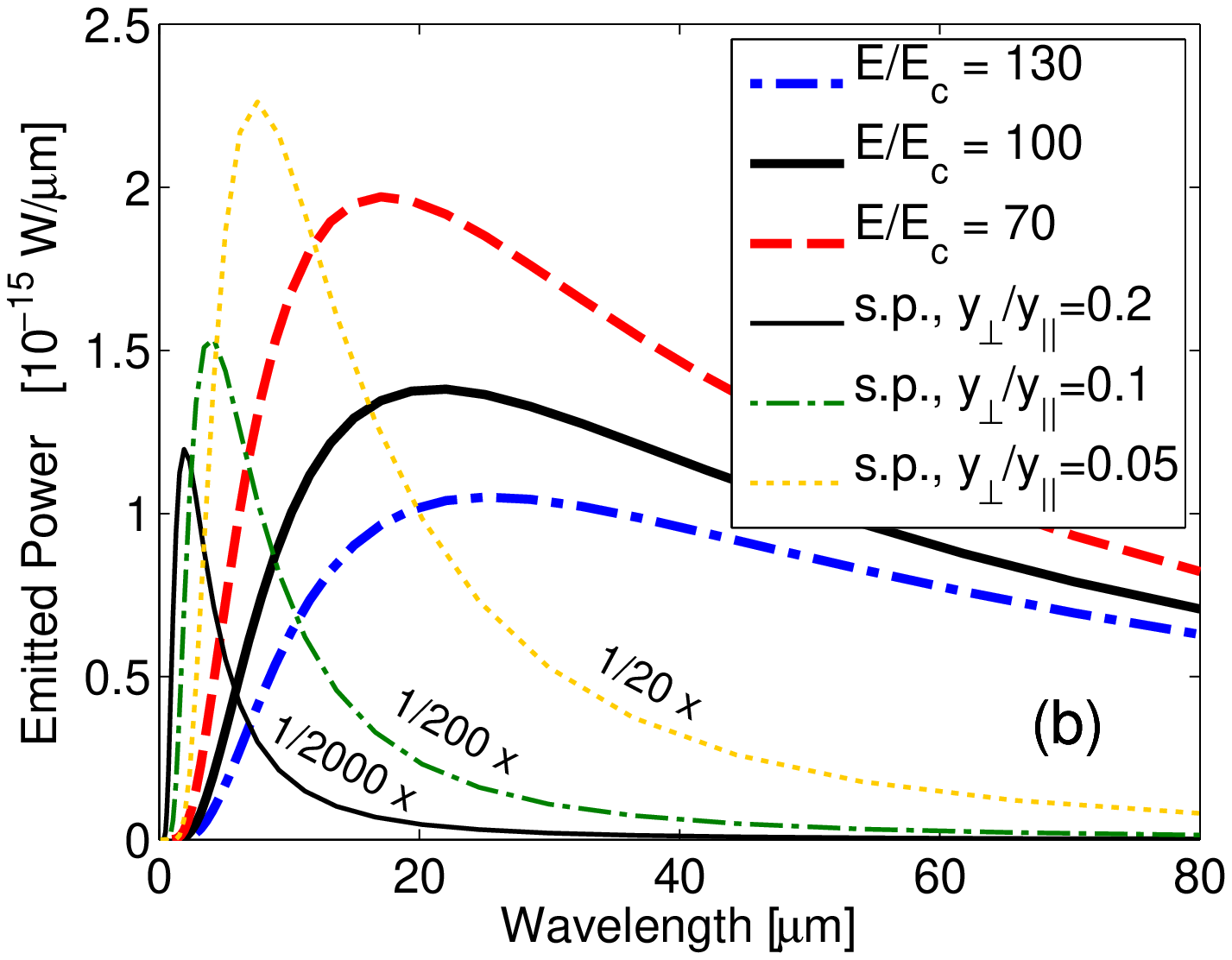}
\caption{(Color online) Synchrotron spectra (average emission per particle) for the runaway
  distributions in (a) Figure \ref{fig:contours7} and (b) Figure
  \ref{fig:contours8}. Emission spectra from the CODE distributions in
  Figures \ref{fig:contours7} and \ref{fig:contours8} are shown in
  solid black, together with spectra from distributions with varying
  electric field strength but otherwise identical physical
  parameters. A magnetic field of $B=3\,$T was used. The synchrotron
  spectra from single particles with $w=50$ and \changed{various pitch angles are also shown.}
(These single-particle spectra are the same in figures a and b, as the particle parameters are independent of simulation settings).
\label{fig:synch_7_8}}
\end{figure}

The per-particle synchrotron spectra generated by the CODE distributions in Figures
\ref{fig:contours7} and \ref{fig:contours8} were calculated using this
formula, and are shown in Figure \ref{fig:synch_7_8}, together with
the spectra radiated by electron distributions for other electric field
strengths. For the physical parameters used, we note that the peak
emission occurs between 7 and 25 $\mu$m. The synchrotron spectra show
a decrease in per-particle emission with increasing electric field strength. Even
though a stronger electric field leads to more particles with high
energy (and thus high average emission), it also leads to a more
narrow distribution in pitch-angle. This reduction in the number of particles with
large pitch-angle leads to a decrease in average emission. Both
figures confirm that the average emission is reduced for higher
electric fields, implying that the latter mechanism has the largest impact on the spectrum.

In calculating the spectra, the runaway region of momentum space, $R$,
was defined such that the maximum particle momentum was
$w_\text{max}=50$ (which translates to $y_\text{max}=500$ and
$y_\text{max}=1000$ respectively for the cases shown in Figures
\ref{fig:contours7} and \ref{fig:contours8}), corresponding to a
maximum particle energy of $\simeq 25$ MeV.  Physically the cutoff at
large energy can be motivated by the finite life-time of the
accelerating electric field and the influence of loss mechanisms such
as radiation. Since the radiated synchrotron power increases with both
particle energy and pitch, this truncation of the distribution is
necessary to avoid infinite emission, although the precise value for
the cutoff depends on the tokamak and on discharge-specific
limitations to the maximum runaway energy. For the low-energy boundary of $R$,
$w_\text{min}=w_c = \left[ (E/E_c)-1 \right]^{-1/2}$ was used, and all
particles with $\xi \in [0,1]$ were included.  Although no
  explicit cutoff was imposed in $\xi$, the distribution decreases
  rapidly as this parameter decreases from 1 (as can be seen in
  Figures \ref{fig:contours7} and \ref{fig:contours8}) and there are
  essentially no particles below some effective cutoff value.

Figure \ref{fig:synch_7_8} also shows the synchrotron spectrum from
single particles with momentum corresponding to the maximum momentum of
the distributions ($w=50$), \changed{and several values of pitch-angle $\ype/\ypa$.}
This single-particle ``approximation'' is
equivalent to using a 2D $\delta$-function model of the
distribution, as was done in Refs. \cite{Finken,Jaspers} (and with
some modification in \cite{Yu}). The figure shows that this
approximation significantly overestimates the synchrotron emission per particle.
Note that in the figure,
the values for the emitted power per particle were divided by a large
number to fit in the same scale. The overestimation is not surprising,
since the $\delta$-function approximation effectively assumes that all particles
emit as much synchrotron radiation as one of the most strongly emitting
particle in the actual distribution. The figure also shows that the
$\delta$-function approximation leads to a different spectrum shape,
with the wavelength of peak emission usually shifted towards shorter wavelengths.
In order to obtain an
accurate runaway synchrotron spectrum, it is thus crucial to use the
full runaway distribution in the calculation.

\begin{figure}
\includegraphics[width=0.8\textwidth]{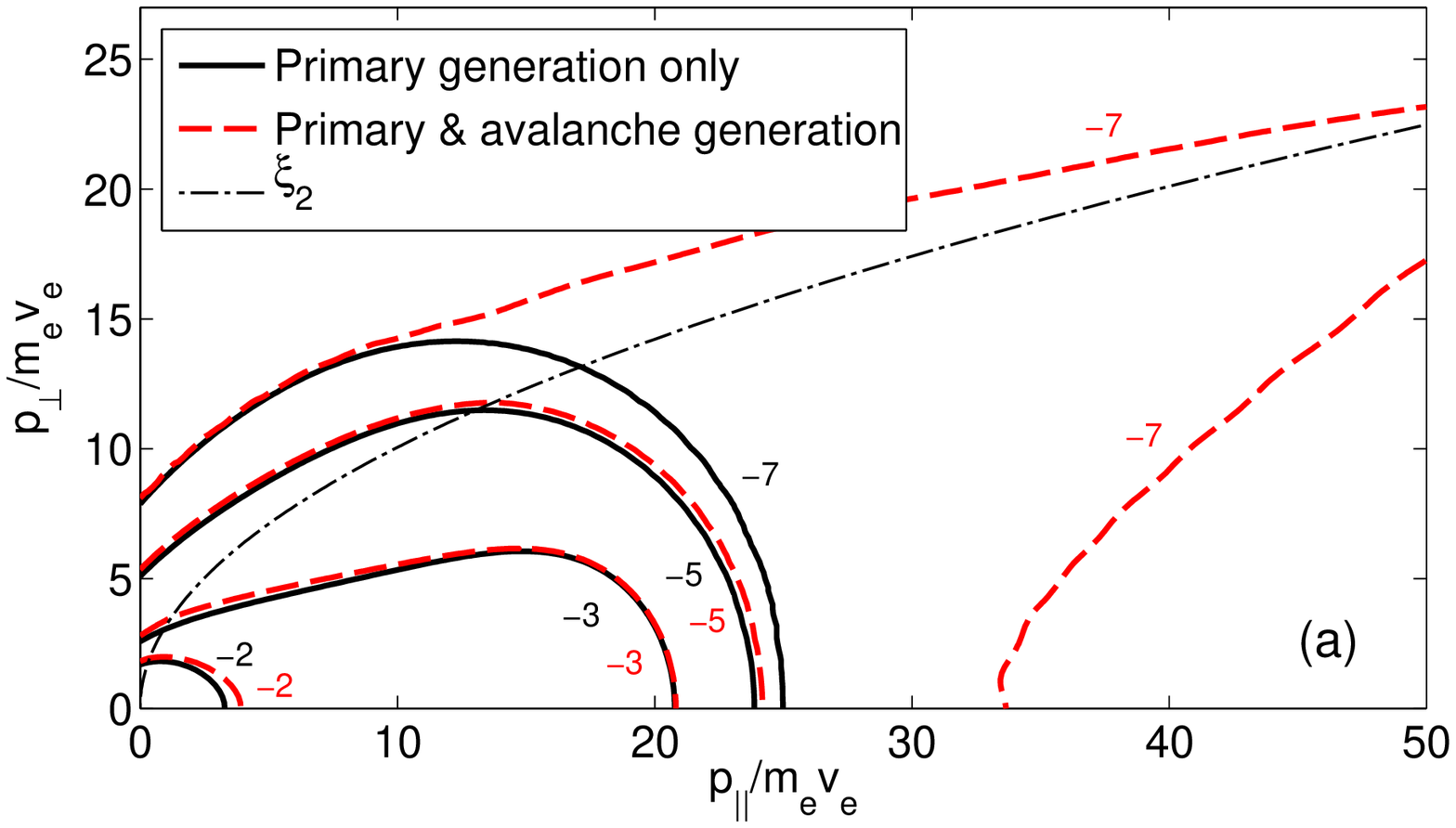}
\includegraphics[width=0.8\textwidth]{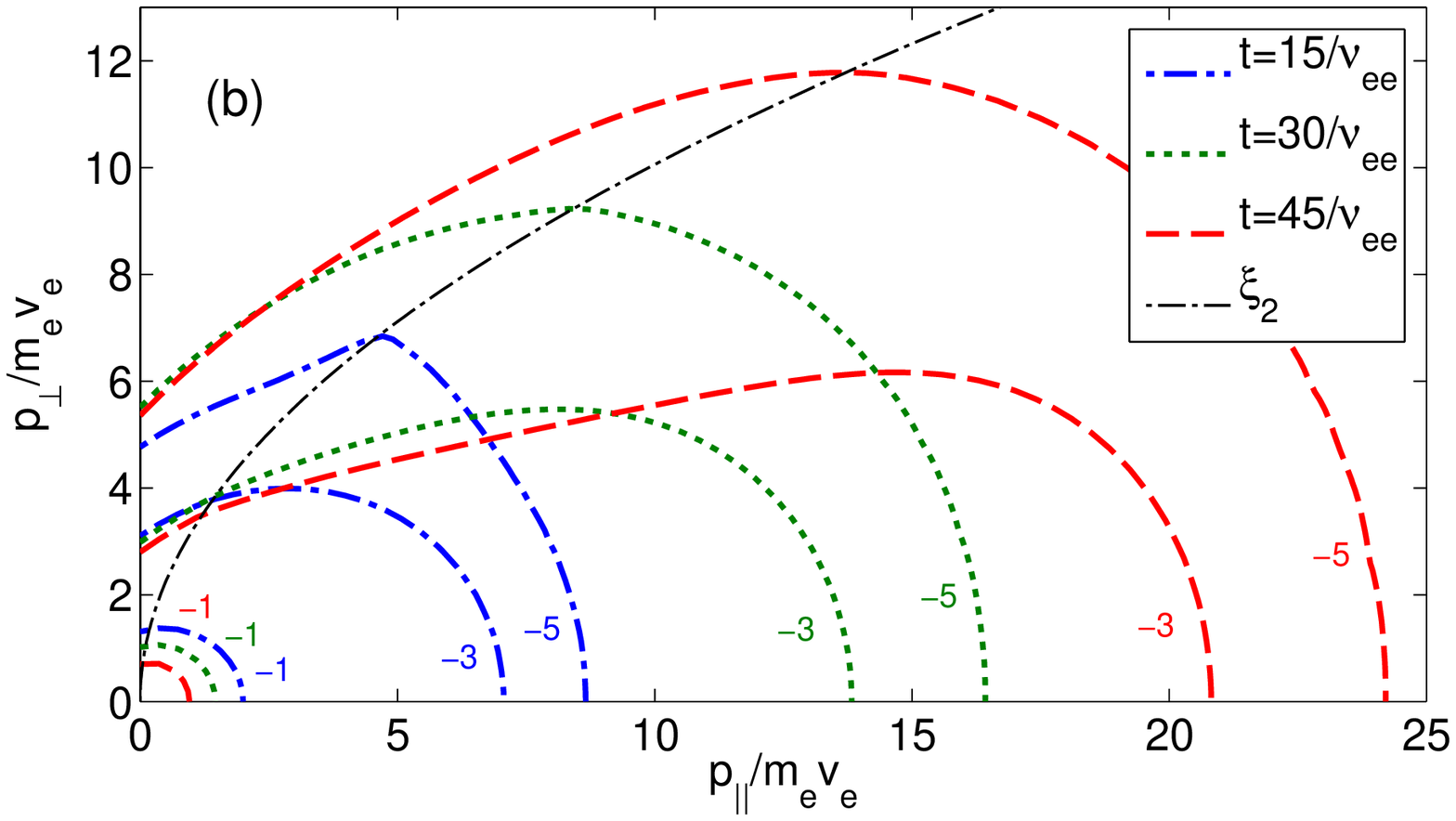}
\caption{\changed{(Color online) (a) Contour plots of the distribution function from primaries only (black solid line), together with a distribution
  obtained with the avalanche source enabled (dashed red line), for $E/E_c=10$  $(\hat{E}=0.523)$, $Z=1$ and $\delta=0.2$ at $t=45/\nu_{ee}$. The quantity plotted is $\log_{10}(F)$. Results are plotted for the numerical parameters $N_y=20$, $y_{max}=100$ and $N_\xi=130$,
  with time step $dt=0.02/\nu_{ee}$.  (b) Contour plots of the distribution function at different times with the avalanche source enabled, using the above parameters. }
  \label{fig:primary}}
\end{figure}

\begin{figure}
\includegraphics[width=0.65\textwidth]{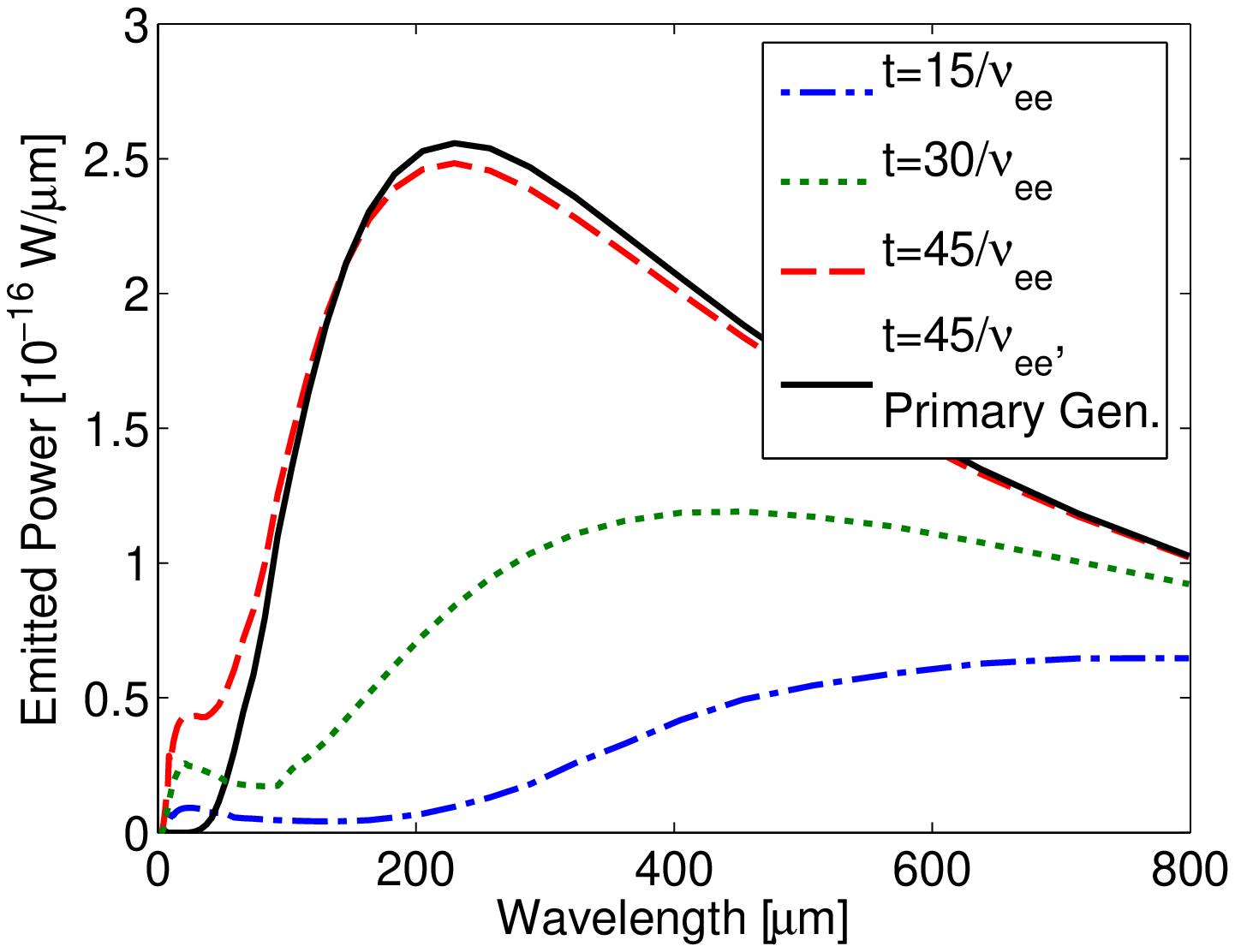}
\caption{\changed{(Color online)
Synchrotron spectra (average emission per particle) for the runaway
  distributions shown in Fig.~\ref{fig:primary} for magnetic field $B=3\,$T.}
  \label{fig:syncprimary}}
\end{figure}

\changed{In the cases shown in Fig.~\ref{fig:synch_7_8}, the
  runaway electron distribution is dominated by secondary
  generation. For comparison, in
  Figs.~\ref{fig:primary}-\ref{fig:syncprimary} we show a case where
  the distribution is dominated by primary generation. Figure
  \ref{fig:primary}a shows contours of a distribution from primaries
  only, together with a distribution obtained with the avalanche
  source enabled, and confirms that the distribution is dominated by primaries, except for a small number of secondary runaways generated along the curve $\xi=\xi_2$. Fig.~\ref{fig:primary}b shows contour
  plots with the avalanche source enabled, for three different times. The
  physical parameters used in Fig.~\ref{fig:primary} are temperature $T=10\;\rm keV$,
  density $n=5\times 10^{19}\;\rm m^{-3}$, effective charge $Z=1$, and
  electric field $E=0.45\;\rm V/m$. The collision time in this case is
  $0.39 \;\rm ms$, so the times shown in the figure correspond to
  $5.9\;\rm ms$, $11.8 \;\rm ms$ and $17.6\;\rm ms$, which correlates
  well with the time-scale of the electric field spike for a typical
  disruption in DIII-D (see e.g. Fig.~2 in \cite{hollmann}.) Figure
  \ref{fig:syncprimary} compares the synchrotron spectra from the
  distributions shown in Fig.~\ref{fig:primary}. The main difference
  compared to the case dominated by secondary generation
  (Fig.~\ref{fig:synch_7_8}) is the generally longer wavelengths in the
  spectrum. The reason is the low runaway electron energy ($w\mylsim
  10$) in the runaway electron distribution in this case. The small peak at short wavelengths in the spectra including the avalanche source stems from the secondary runaways generated at $\xi=\xi_2$ (visible in Fig. \ref{fig:primary}a).}

In principle, we may also use the synchrotron spectra from
distributions calculated through CODE to estimate the maximum energy
of the runaways in existing tokamaks. However, due to the region of
sensitivity of the available detectors, there is only a limited
wavelength range in which calculated spectra can be fitted to
experimental data in order to determine the maximum runaway
energy. The available range often corresponds to the short wavelength
slope of the spectrum, where the emitted power shows an approximately
linear dependence on wavelength.  Indeed, the short-wavelength
spectrum slope has been used to estimate the maximum runaway energy in
experiments \citep{Jaspers}.  If the runaway distribution function is
approximated by a $\delta$-function at the maximum available energy
and pitch angle, there is a monotonic relationship between the
short-wavelength spectrum slope and the maximum particle energy (at
fixed pitch angle).
Such a relationship holds because increasing the particle energy leads
to more emission at shorter wavelengths, resulting in a shift of the
wavelength of peak emission towards shorter wavelengths, and a
corresponding change in the spectrum slope.

Using an integrated synchrotron spectrum from a CODE distribution is
much more accurate than the single particle approximation, but it also
introduces additional parameters ($\hat{E}$, $\delta$, $Z$). If the
physical parameters are well known, a unique relation still holds
between the spectrum slope and the maximum particle energy.  During
disruptions however, many parameters (like the temperature and the
effective charge) are hard to measure with accuracy. As the shape of
the underlying distribution depends on the values of the parameters,
the synchrotron spectrum will do so as well. This complexity is
apparent in Figure \ref{fig:synch_7_8}, where the single particle
approximation produces identical results in the two cases, whereas the
spectra from the complete distributions are widely different. The
dependence on distribution shape makes it possible in principle for
two sets of parameters to produce the same spectrum slope for
different maximum energies. Given this insight, using the complete
runaway distribution when modeling experimentally obtained spectra is
necessary for an accurate analysis and reliable fit of the maximum
particle energy. In this context, CODE is a very useful tool with the
possibility to contribute to the understanding of runaways and their
properties.

\section{Conclusions}
\label{sec:concl}
\changed{
In this work, we have computed the synchrotron emission spectra
from distribution functions of runaway electrons.
The distribution functions are computed efficiently using
the CODE code.}
Both primary (Dreicer) and secondary
(avalanche) generation are included.  A Legendre spectral
discretization is applied to the pitch-angle coordinate, with
high-order finite differences applied to the speed coordinate.  A
nonuniform speed grid allows high resolution of thermal particles at
the same time as a high maximum energy without a prohibitively large
number of grid points.  If secondary generation is unimportant, the
long-time distribution function may be calculated by solving a single
sparse linear system.  The speed of the code makes it feasible to
couple to other codes for integrated modeling of complex processes
such as tokamak disruptions.  CODE has been benchmarked against
previous analytic and numerical results in appropriate limits, showing
excellent agreement.
\changed{In the limit of strong avalanching, CODE demonstrates
agreement with the analytic distribution function (\ref{eq:anal_ava})
from Ref. \onlinecite{Fulop2006}.}

\changed{
The synchrotron radiation spectra
are computed by convolving the distribution function
with the single-particle emission.}
We find that the
radiation spectrum from a single electron at the maximum energy
can
differ substantially from the overall spectrum generated by a
distribution of electrons.
Therefore, experimental estimates of maximum runaway energy
based on the single-particle synchrotron spectrum are likely to be inaccurate.
A detailed study of the distribution-integrated synchrotron spectrum
and its dependence on physical parameters
can be found in Ref. \cite{Adam}.

In providing the electron distribution functions (and thus knowledge
of a variety of quantities through its moments), the applicability of
CODE is wide, and the potential in coupling CODE to other software,
e.g. for modeling of runaway dynamics in disruptions, is promising.
For a proper description of the runaways generated in disruptions it is
important to take into account the evolution of the radial profiles of
the electric field and fast electron current self-consistently. This
can be done by codes such as GO, initially described in Ref.
\cite{smith06go} and developed further in
Refs.~\cite{gal08runaway,feher11simulation}. GO solves the equation
describing the resistive diffusion of the electric field in a
cylindrical approximation coupled to the runaway generation rates. In
the present version of GO, the runaway rate is computed by approximate
analytical formulas for the primary and secondary generation. Using
CODE, the analytical formulas can be replaced by a numerical solution
for the runaway rate which would have several advantages. One advantage would be
that Dreicer, hot-tail and secondary runaways could all be calculated
with the same tool, avoiding the possibilities for double-counting and
difficulties with interpretations of the results. Also, in the present
version of GO, it is assumed that all the runaway electrons travel at
the speed of light, an approximation that can be easily relaxed using
CODE, which calculates the electron distribution in both energy and
pitch-angle. Most importantly, the validity region of the results
would be expanded, as the analytical formulas are derived using
various assumptions which are often violated in realistic
situations. The output would be a self-consistent time and space
evolution of electric field and runaway current, together with the
electron distribution function. This information can then be used for
calculating quantities that depend on the distribution function,
such as the synchrotron emission or the kinetic instabilities driven by
the velocity anisotropy of the runaways.

\begin{acknowledgments}
  This work was supported by US Department of Energy grants DE-FG02-91ER-54109 and DE-FG02-93ER-54197, by
  the
  Fusion Energy Postdoctoral Research Program administered by the Oak
  Ridge Institute for Science and Education, and by the European
  Communities under Association Contract between EURATOM and {\em
    Vetenskapsr{\aa}det}. The views and opinions expressed herein do
  not necessarily reflect those of the European Commission. The
  authors are grateful to J. Ryd\'en, G. Cs\'ep\'any, G. Papp, P. Helander, \changed{E. Nilsson}, J. Decker, and Y. Peysson for fruitful
  discussions.
\end{acknowledgments}

\appendix

\section{Integrals of Legendre Polynomials}
\label{a:Legendres}

Here we list several identities for Legendre polynomials
which are required for the spectral pitch-angle discretization.
To evaluate the $\xi$ integral of the $\xi\; \partial F/\partial y$ term in (\ref{eq:normalizedKE}), we use the recursion relation
\begin{equation}
\xi\; P_L(\xi) = \frac{L+1}{2L+1} P_{L+1}(\xi) + \frac{L}{2L+1} P_{L-1}(\xi)
\end{equation}
where $P_{L-1}$ is replaced by 0 when $L=0$. Applied to the relevant integral in (\ref{eq:normalizedKE}),
and noting the orthogonality relation
$(2L+1)2^{-1}\int_{-1}^1 P_L(\xi) P_\ell(\xi)d\xi = \delta_{L,\ell}$,
we find
\begin{equation}
\frac{2L+1}{2}\int_{-1}^1 d\xi\; \xi\;  P_L(\xi) P_\ell(\xi) =
\frac{L+1}{2L+3}\delta_{\ell,L+1} + \frac{L}{2L-1} \delta_{\ell,L-1}.
\end{equation}
Similarly, to evaluate the $\xi$ integral of the $\partial F/\partial\xi$ term in
(\ref{eq:normalizedKE}), we use the recursion relation
\begin{equation}
(1-\xi^2)(dP_L/d\xi) = LP_{L-1}(\xi) - L \xi P_L(\xi)
\end{equation}
to obtain
\begin{equation}
\frac{2L+1}{2}\int_{-1}^1 d\xi\; P_L(\xi) (1-\xi^2)\frac{dP_\ell}{d\xi}
= \frac{(L+1)(L+2)}{2L+3}\delta_{\ell,L+1} - \frac{(L-1)L}{2L-1} \delta_{\ell,L-1}.
\end{equation}
Finally, the pitch-angle scattering collision term gives the integral
\begin{equation}
\frac{2L+1}{2}\int_{-1}^1 d\xi\; P_L(\xi) \frac{\partial}{\partial \xi} (1-\xi^2) \frac{\partial}{\partial \xi} P_\ell(\xi)
= -(L+1)L\delta_{L,\ell}.
\end{equation}

\bibliography{CODE}

\end{document}